\documentclass[
]{ceurart}

\usepackage{listings}
\usepackage{booktabs}
\usepackage{hyperref}

\begin{document}

\copyrightyear{2024}
\copyrightclause{Copyright for this paper by its authors.  Use permitted under Creative Commons License Attribution 4.0 International (CC BY 4.0).}

\conference{NORMalize 2024: The Second Workshop on the Normative Design and Evaluation of Recommender Systems, October 18, 2024, co-located with the ACM Conference on Recommender Systems 2024 (RecSys 2024), Bari, Italy}
\title{Value Identification in Multistakeholder Recommender Systems for Humanities and Historical Research:\\The Case of the Digital Archive Monasterium.net}

\author[1,2]{Florian Atzenhofer-Baumgartner}[%
orcid=0000-0001-8157-8629,
email=atzenhofer@acm.org,
url=https://atzenhofer.github.io]
\cormark[1]
\address[1]{Institute of Interactive Systems and Data Science, Graz University of Technology, Graz, Austria}
\address[2]{Department of Digital Humanities, University of Graz, Graz, Austria}

\author[3,4]{Bernhard C. Geiger}[%
orcid=0000-0003-3257-743X,
email=geiger@ieee.org
]
\fnmark[1]
\address[3]{Know Center Research GmbH, Graz, Austria}
\address[4]{Signal Processing and Speech Communication Laboratory, Graz University of Technology, Graz, Austria}

\author[4]{Georg Vogeler}[%
orcid=0000-0002-1726-1712,
email=georg.vogeler@uni-graz.at
]
\fnmark[1]

\author[1,3]{Dominik Kowald}[%
orcid=0000-0003-3230-6234,
email=dkowald@know-center.at
]
\fnmark[1]

\cortext[1]{Corresponding author.}
\fntext[1]{These authors contributed equally.}

\begin{abstract}
Recommender systems remain underutilized in humanities and historical research, despite their potential to enhance the discovery of cultural records. This paper offers an initial value identification of the multiple stakeholders that might be impacted by recommendations in  Monasterium.net, a digital archive for historical legal documents. Specifically, we discuss the diverse values and objectives of its stakeholders, such as editors, aggregators, platform owners, researchers, publishers, and funding agencies. These in-depth insights into the potentially conflicting values of stakeholder groups allow designing and adapting recommender systems to enhance their usefulness for humanities and historical research. Additionally, our findings will support deeper engagement with additional stakeholders to refine value models and evaluation metrics for recommender systems in the given domains. Our conclusions are embedded in and applicable to other digital archives and a broader cultural heritage context. 
\end{abstract}



\maketitle

\section{Introduction}

Recommender systems (RecSys) analyze past usage behavior to suggest potential relevant content to users~\cite{jannach2010recommender}. Although RecSys have been applied in many domains, such as music~\cite{kowald2020unfairness,Schedl2021}, movies~\cite{harper2015movielens,kowald2022popularity}, or online marketplaces~\cite{lacic2014towards}, the fields of cultural heritage and digital humanities (DH) have, so far, only received minimal and narrowly focused attention in RecSys research. 

However, a potentially important use case for RecSys lies in the aggregation and dissemination of cultural heritage objects and primary sources from archives. This is particularly relevant for disciplines such as diplomatics and other auxiliary sciences, where scholars frequently work with historical legal documents, including so-called charters (see Section~\ref{Items} for more details on charters as the main item type for RecSys in DH and historical research).

It can be estimated that there exist millions of charters just in Central Europe, with the majority of them not yet digitized. There are several attempts to make charters digitally available, which are usually part of general archival digitization attempts and dedicated online publications, such as with Cartae Europae Medii Aevi (CEMA)~\cite{Perreaux_2021} or the Digitale Charterbank Nederland (DCN)~\cite{Burgers_Hoekstra_2021}. The largest of them is Monasterium.net\footnote{\url{https://www.monasterium.net/mom/home}.}. 

\vspace{2mm} \noindent \textbf{The digital archive Monasterium.net} currently covers more than 650,000 charters, and has become a vital resource for studies across many fields beyond diplomatics, including paleography, art history, and linguistics. It attracts an international user base with approximately 4,000 monthly visits, primarily from Europe and North America. A core group of about 3,000 subscribed users has access to data creation/annotation functionalities, and it includes university students, amateur historians, and scholars of history and adjacent disciplines. Monasterium.net currently relies on string-matching-based full-text search with additional drill-down options. While this approach has served users thus far, it faces significant limitations when dealing with the platform's high-dimensional and varied data. The current search mechanism struggles to effectively filter and rank the vast amount of information, leading to suboptimal retrieval of relevant documents. This inefficiency is particularly problematic given the platform's extensive document collection, which continues to grow through ongoing digitization efforts.

Recognizing these challenges, Monasterium.net is undergoing a substantial redevelopment through the ERC project 'From Digital to Distant Diplomatics' (DiDip), which aims to integrate machine learning pipelines to enhance the user experience in finding and analyzing relevant materials~\cite{Luger_Nicolaou_Decker_Atzenhofer-Baumgartner_Lamminger_Vogeler_Aoun_Kovács_2023}. The implementation of RecSys, also driven by awareness and consideration of the platform's stakeholder interests, holds significant promise for addressing said limitations. 

\vspace{2mm} \noindent \textbf{Employing RecSys for humanities and historical research} aligns with the broader trend and growing need in cultural heritage venues of managing and interpreting increasingly large and complex datasets. Despite these potential benefits and the ongoing efforts to implement AI-driven solutions in cultural heritage portals, the use of RecSys in this sector remains under-explored, particularly in the context of historical research and digital archives. While AI applications are being implemented, scrutinized, and debated in carious cultural heritage contexts~\cite{Colavizza_Blanke_Jeurgens_Noordegraaf_2022,Cushing_Osti_2023,Lee_2023}, RecSys are widely overlooked, with relevant discussions at best emerging under the guise of ``discovery systems''.

RecSys have been shallowly explored in cultural heritage contexts, their application has primarily focused on physical museums and exhibition spaces \cite{Pavlidis_2019, Casillo_Colace_Conte_Lombardi_Santaniello_Valentino_2023}, often leveraging visitor location data and content-based approaches, fuelled by standardized ontologies. However, the unique challenges posed by digital archives (re-)publishing historical documents have received little attention. Unlike systems dealing with physical curations and visitor engagement, digital archives like Monasterium.net, which are significantly more research-oriented, arguably pose different requirements. Besides, multistakeholder and value-oriented aspects have been neglected in both. Our work aims to address this gap by examining and paving the way for specific needs and values of stakeholders in the given domain, and by this extending the discourse on RecSys in cultural heritage beyond the traditional museum-centric approach.

We argue that the nature of cultural records as items, such as charters, and the unique needs of scholars as users, such as historians, introduce distinct challenges and normative considerations that must be addressed for an effective RecSys implementation in the fields of humanities and historical research. This coincides with the call of the RecSys community for a deeper exploration of concrete use cases and multistakeholder settings in the context of normative and fair AI~\cite{Jannach_Abdollahpouri_2023, Deldjoo_Jannach_Bellogin_Difonzo_Zanzonelli_2024}. Also, it highlights the importance of value-aware RecSys~\cite{Abdollahpouri_Adomavicius_Burke_Guy_Jannach_Kamishima_Krasnodebski_Pizzato_2020} and the consideration of potential conflicts in multi-objective settings~\cite{Jannach_Abdollahpouri_2023} in light of a yet under-researched domain. 

\vspace{2mm} \noindent \textbf{The present paper} thus provides a first analysis of how items, users, and corresponding values could be modeled in the context of RecSys for digital archives, focusing specifically on Monasterium.net. Before analyzing the stakeholders and their values in this setting (Section~\ref{Stakeholders}), we first give a more detailed description of charters as the main item type investigated in RecSys for humanities and historical research (Section~\ref{Items}). We conclude the paper by discussing some limitations in generalizing our findings and outlining plans for future research.

\section{Charters as the Main Item Type for RecSys in Monasterium.net}\label{Items}

Charters are documents that record legal actions, issued and authenticated according to specific formal requirements. They document transactions or agreements regarding land, property, privileges, or legal rights between two or more parties. As such, they are arguably the information-richest source for researching the past: they allow insights into the legal, social, economic, and cultural aspects of former societies~\cite{Slavin_2012}. These documents have survived in various forms, either as originals or copies, and can be distinguished by their legal effect or administrative purpose. Copies may serve different functions, such as duplication, transcription, or translation, and vary in their credibility and means of authentication.

\begin{figure}[!h]
  \centering
  \includegraphics[scale=0.8]{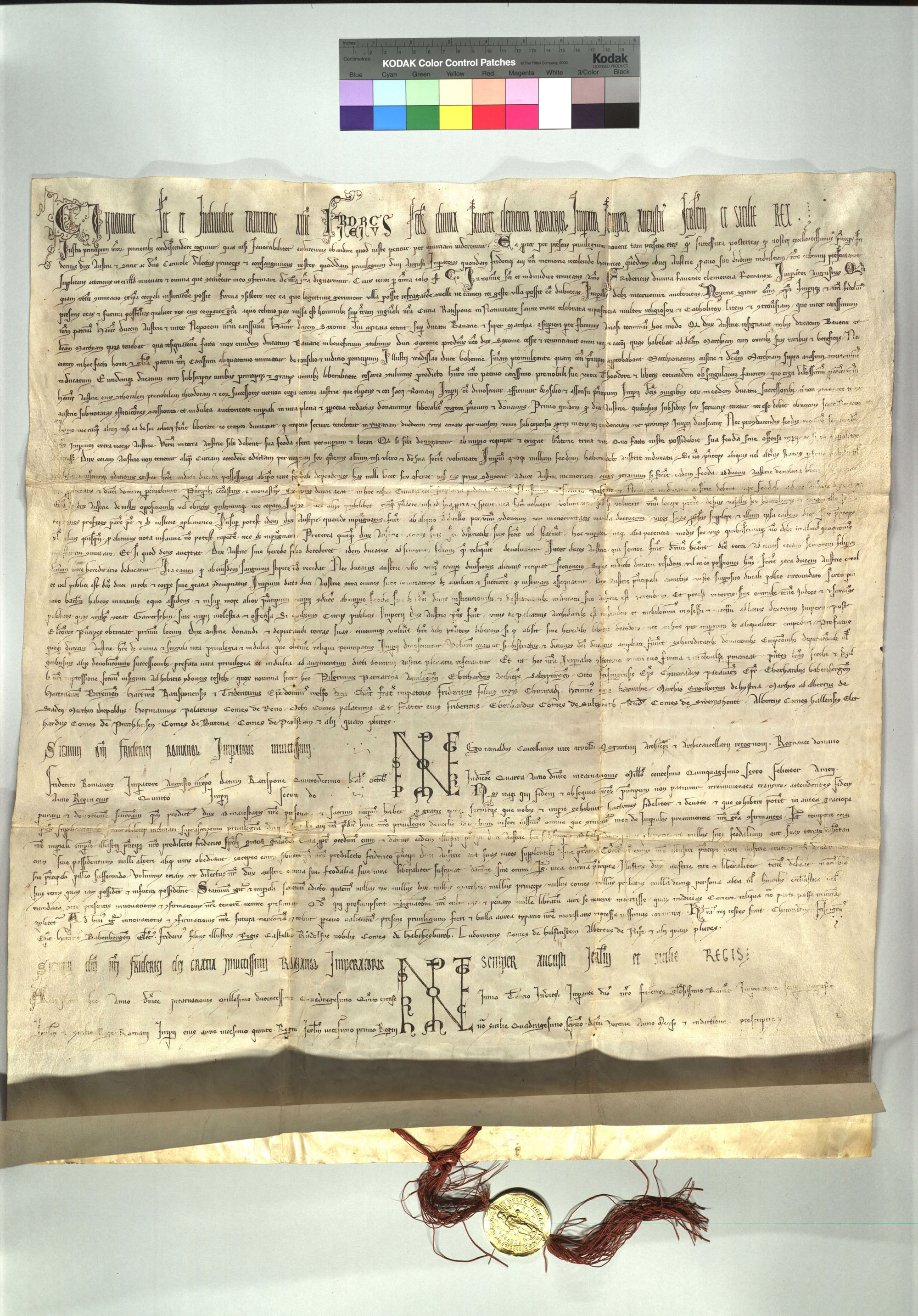}
  \caption{Example document: Recto side of a charter by Holy Roman Emperor Frederick II to Babenberg Duke Frederick II containing the text of the Privilegium maius - itself a forged adaptation of the Privilegium minus by Frederick I -, confirming it, and extending its terms. Purportedly issued in Verona, Italy in June 1245, however forged in 1358/59. Its illegitimacy, identified by contemporaries and proven centuries later, had considerable historical and legal implications. Adapted from \url{https://www.archivinformationssystem.at/detail.aspx?id=183092}, re-published in \url{https://www.monasterium.net/mom/AT-HHStA/AUR/AUR_1245_VI/charter}.}
\end{figure}

On Monasterium.net, charters are either digitized as scans of manuscripts or scholarly editions, complemented by standardized metadata, and available as semi-structured data, such as Extensible Markup Language (XML). Specialized schemas, such as the Charter Encoding Initiative (CEI)~\cite{Vogeler_2005} which builds on existing diplomatic conventions~\cite{Cárcel_Ortí_1997}, enhance their accessibility and facilitate advanced analysis. Besides image scans, transcriptions, and abstracts, the metadata can hold information on the material (e.g., parchment), production process (e.g., issuer, place, receiver), means of authentication (e.g., seals, notarial signs), and meta-commentary (e.g., formal diplomatic analysis). Given this high dimensionality, metadata is often sparsely populated, since data and metadata have uneven rates of transmission and availability.

Historically, charters were stored and collected pragmatically. Today, they are primarily organized according to archival principles mandated by laws or guidelines. Beyond archival mandates, charters are also gathered based on specific scholarly interests, leading to curations often grouped by some measure of similarity or ordered by a taxonomy. Examples of this on Monasterium.net are given with critical (micro-)editions~\cite{Ambrosio_2020} or the flagship collection named ``illuminated charters''~\cite{Gneiß_Zajic_2015,Bartz_Gneiß_2019}. Advances in archival principles may result in extensions of the governing metadata schema, enabling fine-grained annotation, such as regarding entities or style~\cite{Winslow_2020}.

Monasterium.net stands out among charter platforms due to its broad definition of charters and thus their exceptionally large variety, including diplomas, deeds, privileges, mandates, orders, and letters. Additionally, it primarily hosts charters that go beyond high political ranks, like royal or papal issuers. These so-called private charters are especially valuable for constructing comprehensive historical narratives rather than focusing on singularities. Covering an extensive period from the early middle ages to the early modern era and beyond (ca. 700-1800 AD), the platform has a geographical focus on (central) Europe. Consequently, it contains and supports heavily multilingual data, ranging from modern English to Old Church Slavonic \cite{Kovács_Atzenhofer-Baumgartner_Aoun_Nicolaou_Luger_Decker_Lamminger_Vogeler_2022}.

\section{An Analysis of Monasterium.net Stakeholders and their Values}\label{Stakeholders}

The present conceptualization and terminology of stakeholders is adapted from respective (evaluation) surveys~\cite{Abdollahpouri_Adomavicius_Burke_Guy_Jannach_Kamishima_Krasnodebski_Pizzato_2020, Zangerle_Bauer_2023} and informed by a comprehensive literature review, contributions to the platform's development and conceptualization, and contact with its stakeholders. While this work can be generalized to other digital archives, the main focus is put on Monasterium.net and its extensive collection of charters.

\begin{table}[!h]
\centering
\caption{Key example objectives and values in a multistakeholder recommender system for Monasterium.net.}
\begin{tabular}{l l l}
\toprule
\textbf{Category} & \textbf{Example} & \textbf{Objectives and Values} \\
\midrule
\textbf{Upstream} & Editors & 
\begin{tabular}[c]{@{}l@{}} 
Ensure high-quality curations \\ Impact scholarly work through recommendations \\ Achieve visibility and prestige
\end{tabular} \\
\midrule
\textbf{Providers} & Aggregators & 
\begin{tabular}[c]{@{}l@{}} 
Provide accurate and comprehensive data \\ Facilitate efficient data access \\ Become leading aggregator
\end{tabular} \\
\midrule
\textbf{System} & Platform owners & 
\begin{tabular}[c]{@{}l@{}} 
Increase user satisfaction with recommendations \\ Drive strategic growth of the platform \\ Foster collaboration with stakeholders
\end{tabular} \\
\midrule
\textbf{Consumers} & Researchers & 
\begin{tabular}[c]{@{}l@{}} 
Obtain relevant recommendations \\ Rely on accurate and useful results \\ Generate new insights and research questions
\end{tabular} \\
\midrule
\textbf{Downstream} & Publishers & 
\begin{tabular}[c]{@{}l@{}} 
Increase novelty of published works \\ Publish accurate and impactful research \\ Generate revenue through publications
\end{tabular} \\
\midrule
\textbf{Third Parties} & Funding agencies & 
\begin{tabular}[c]{@{}l@{}} 
Support sustainability and accessibility of data \\ Promote extensive use of funded curations \\ Ensure maximum impact of funded projects
\end{tabular} \\
\bottomrule
\end{tabular}
\label{tab:stakeholder_summary}
\end{table}

\subsection{Upstream Stakeholders}
Upstream stakeholders are potentially impacted by recommendations, but are not directly providing the items. Thus, they play a major role in the preservation, enrichment, and dissemination of charters.

\vspace{2mm} \noindent \textbf{Content creators} include archivists, scholars, editors, and curators. Their primary goals are to ensure high-quality curations as research objects, to achieve visibility and, with that, prestige within their fields. Monasterium.net is inherently collaborative, allowing verified users to engage with data extension and enrichment, and to create interpretive copies: it represents a type of volunteer-based crowdsourcing~\cite{Andro_Saleh_2017}. This involves producing and compiling collections that extend beyond typical archival organization and description, which further promotes its reuse in study and teaching. Content creators highly value the impact and reach of their curations on other scholarly productions. 

\vspace{2mm} \noindent \textbf{Historical document hosts}, such as archives, libraries, museums, and private hosts, are responsible for preserving and providing (analog) access to historical documents. They differ slightly in what they value as well as in the extent to which they are open to the general public and the ratio between accessible and yet non-accessible data: the digitization status directly affects document accessibility and usability.\footnote{For instance, some Lower Austrian monasteries' decision to digitize their collections, driven by their challenges in managing physical access, marked the origin of Monasterium.net. This initiative not only addressed their access issues but also set a foundation for pilot digitization strategies in other (small) archives~\cite{Aigner_2002}.} Archives of all sizes aim to have their collections represented fairly, despite differences in quality and quantity of metadata. They strive to make resources accessible and visible, even when data is incomplete or partially lost. Besides, hosts value the effectiveness of their collections' access and use. This reflects in how well their material is referenced or re-published through aggregators and later in scholarly works. As such, they utilize Monasterium.net as a multiplier to promote their holdings.

\subsection{Provider Stakeholders}
Providers are key to ensuring an improved (digital) accessibility and representation of cultural records.

\vspace{2mm} \noindent \textbf{Data Providers} include operators of archival information systems, aggregators, and cataloging services. They are standalone entities that offer application programming interfaces or even full-stack solutions to integrate archives more effectively. Thus, they represent proxies between archives and re-publishing platforms. Usually, their focus is laid on data accuracy, completeness, and recency. As gatekeepers for user feedback, they can help ensure high-quality data representation, which is crucial since aggregation might omit metadata~\cite{Tóth-Czifra_2020}. Data providers must balance the needs of archives for accuracy and completeness with the demands of re-publishers, which may require adapting to different formats. With a commercial and economy-of-scale mindset, they often start by serving a few archives and expand to many, potentially making archives dependent on their services. They often aim to become leading aggregators of digital objects, and thus a main beneficiary of successful recommendations.

\vspace{2mm} \noindent \textbf{Digitization services} are usually performed by small teams and within partnerships between companies, institutions, and other entities. These stakeholders are crucial for converting physical documents into digital formats to make them usable either for hosts or providers directly. The primary goals of digitization services involve ensuring high-quality digitization, efficiency in processing, and the accuracy of the digital representations as per some standard. Their contributions are fundamental for all other stakeholders. Consequently, they value the impact of digitization quality on recommendations and the reception of their work.

\subsection{System Stakeholders}
System stakeholders are crucial for the management, integrity, and maintenance of the platform itself.

\vspace{2mm} \noindent \textbf{Platform owners}, such as the International Centre for Archival Research (ICARus) for Monasterium.net, are responsible for the platform’s operation and advancement. Their goals include ensuring user satisfaction and decent engagement metrics. They balance diverse stakeholder needs, negotiate contracts, form strategic partnerships to expand data coverage, and manage relationships to enhance the platform’s influence and effectiveness. Therefore, they value strategic growth, stakeholder collaboration, and platform impact.

\vspace{2mm} \noindent \textbf{Moderators}, often subject-matter experts, consult on and authenticate changes to existing digital objects. As creators, researchers, and users of the data, they ensure document integrity and authenticity on Monasterium.net. Their expertise is critical for managing the collaborative aspect of the platform, as they ensure that contributions and modifications meet certain standards. In a RecSys, their interest besides data integrity and quality lies in the influence of moderation and review on recommendations.

\vspace{2mm} \noindent \textbf{Developers}, such as software engineers, web developers, and data analysts, are responsible for creating and maintaining the technology underlying the RecSys. They focus on designing, coding, testing, and deploying the software stack that supports the platform. Key objectives for developers include optimizing recommendation algorithms for relevance and accuracy, ensuring high performance and scalability, as well as enabling effective analytics to improve the system's capabilities. Along with administrators, they must also ensure that the platform manages the integration of new data without degrading the quality of recommendations or compromising system performance.

\subsection{Consumer Stakeholders}
Consumers directly receive the recommendations by the system. Each ``consumer'' stakeholder group interacts with the platform in different ways and at different levels of intensity.

\vspace{2mm} \noindent \textbf{Researchers} use a digital archive primarily to find relevant documents and generate new insights. Historians, in particular, are focused on analyzing documents and uncovering truths to form accurate, novel, and convincing historical narratives. They aim to compile and compare sources, validate authenticity, possibly identify forgeries, and provide a thorough understanding of historical events by using established methodologies to assess and interpret documents, such as the diplomatic method. They often do so by working over various periods of time, with sub-tasks taking hours to days and overarching tasks possibly taking months, which implies that scholars have short-term and long-term goals. Accordingly, researchers value the ability to search and retrieve information efficiently, and to rely on the relevance and accuracy of recommendations. For historians, this might manifest in (unforeseen) inspiration to formulate new research questions or generate hypotheses and complex historical queries. Additionally, session-based evaluation might pose valuable insights into how historians work.

\vspace{2mm} \noindent \textbf{Educators}, such as professors and teachers, use Monasterium.net to inform their teaching materials and engage students with primary sources. Their goals include integrating historical documents into curricula, teaching critical thinking, and providing students with hands-on experience in analyzing historical texts~\cite{Ambrosio_2011}. Accordingly, \textbf{students} interact with historical document platforms to explore and engage with historical content as part of their learning process. This means that both stakeholders benefit from recommendations to draw learners into the material and sustain their interest, likely reflecting in the time spent with individual items or overall on the platform.

\vspace{2mm} \noindent The \textbf{general public} typically uses digital archives for general or personal sporadic interest. Their goals may include casual exploration of historical documents, learning about historical events, or satisfying personal curiosity. They value the ability to easily access and navigate content, as well as to discover information that aligns with their individual interests. 

\subsection{Downstream Stakeholders}
Downstream stakeholders are impacted by the choices of recommendation consumers, but do not directly receive recommendations.

\vspace{2mm} \noindent \textbf{Publishers}, including those of journals and books, focus on the accuracy and novelty of charters as data as well as the impact and reach of their publications. Their goal is to publish works that contribute new insights or noteworthy advances in the field and, to some extent, generate revenue. Effective recommendations and new document curations emerging from the works of consumers or content creators, such as scholars and editors, helps publishers achieve this goal. 

\vspace{2mm} \noindent \textbf{Educational platforms and media} include learning environments and journalistic coverage. Albeit a relatively small group, these stakeholders play a crucial role in disseminating historical knowledge and engaging the public. Similar to educators, their goals include telling compelling stories while offering accurate historical context to historical documents. Most importantly, they aim to generate public value.

\subsection{Third-Party Stakeholders}
Finally, third-party stakeholders do not directly interact with the platform itself, but are impacted by its recommendations and interactions.

\vspace{2mm} \noindent \textbf{Funding agencies} include governments, educational institutions, foundations, and companies that provide financial support for the digitization and preservation of charters. They aim to support educational and scholarly activities, ensure the sustainability and accessibility of cultural heritage, and promote the use of such documents in research and teaching. Funding agencies value the extent of digitized collections and their (re-)use in academic research and education.

\vspace{2mm} \noindent \textbf{Policymakers}, including cultural heritage agencies and legislators, shape policies related to the preservation, access, and use of historical documents. Their goals are to develop effective cultural heritage policies, ensure the preservation of historical records, and promote fair and sustainable access to historical information. Collaborating with this group is crucial for addressing challenges such as legal constraints, licensing, and usage rights, which affect the utility of these information sources to scholars and thus the utility of recommendations~\cite{Nix_Decker_2023}.

The implementation of a RecSys in this context must consider the interplay of said stakeholder interests. Particularly crucial are potential conflicts with long-lasting negative effects, such as researchers acting as content creators, potentially influencing both input and evaluation of recommendations; platform owners focused on growth, who might prioritize increasing document count and site traffic over balanced archival representation; and competing archives seeking greater visibility for their collections, which could lead to increased bias in recommendations. These example scenarios underscore the need for thoughtful system design that balances diverse stakeholder needs while maintaining fairness and transparency.

\section{Conclusion and Future Work}

Integrating RecSys into DH and digital archives presents both opportunities and challenges. Unlike commercial RecSys that can easily use revenue as a proxy for stakeholder value, those in DH must be designed to navigate large and complex datasets in mostly non-commercial settings to enhance access to and understanding of the human cultural record. Moving forward, key challenges include prioritizing the process of gathering scientific insights with effective recommendations, while not neglecting other stakeholders. While the potential of RecSys is quite clear in how they can filter and distill information for scholars, their many facets of utility remain difficult to quantify.

To tackle these challenges, a traditionally different and more nuanced approach to user and values modeling is necessary. A limitation of our current paper is that our analysis was informed by only a small selection of experts who are heavily involved in the domain. This insider perspective, though insightful, does not yet consider other stakeholders sufficiently to derive a full set of values and metrics.

\vspace{2mm} \noindent \textbf{Future work.} In the future, we aim to collaborate more closely with other stakeholders to derive more concrete values and potential metrics through structured interviews and user studies, following best practices~\cite{Smith_Satwani_Burke_Fiesler_2024}. This work represents a first step in refining RecSys to better serve both expert scholars and the general public. Platforms like Monasterium.net offer a glimpse of the potential of such systems in meeting the diverse objectives prevalent in the nexus of humanities and historical research. While our analysis focuses on a specific digital archive, the stakeholder landscape and value considerations identified here likely extend to other cultural heritage institutions and potentially to RecSys applications in knowledge-focused domains more broadly.

\vspace{2mm} \noindent \textbf{Acknowledgements}. 
The work presented in this paper has been supported by the ERC Advanced Grant project (101019327) ``From Digital to Distant Diplomatics''. Additionally, this research was supported by the Know Center Research GmbH within the COMET — Competence Centers for Excellent Technologies Programme, funded by bmvit, bmdw, FFG, and SFG. 

\bibliography{bibliography}

\end{document}